# COMPARING REPOSITORY TYPES.

## Challenges and barriers for subject-based repositories, research repositories, national repository systems and institutional repositories in serving scholarly communication

Chris Armbruster, Max Planck Digital Library, Max Planck Society
Invalidenstrasse 35, D-10115 Berlin - www.mpdl.mpg.de
Executive Director, Research Network 1989 - www.cee-socialscience.net/1989

Laurent Romary, INRIA-Gemo & Humboldt Universität zu Berlin (Institut für Deutsche Sprache und Linguistik)
Dorotheenstrasse 24, D-10117 Berlin - www.linguistik.hu-berlin.de/

### Abstract
After two decades of repository development, some conclusions may be drawn as to which type of repository and what kind of service best supports digital scholarly communication, and thus the production of new knowledge. Four types of publication repository may be distinguished, namely the subject-based repository, research repository, national repository system and institutional repository.

Two important shifts in the role of repositories may be noted. With regard to content, a well-defined and high quality corpus is essential. This implies that repository services are likely to be most successful when constructed with the user and reader uppermost in mind. With regard to service, high value to specific scholarly communities is essential. This implies that repositories are likely to be most useful to scholars when they offer dedicated services supporting the production of new knowledge.

Along these lines, challenges and barriers to repository development may be identified in three key dimensions: a) identification and deposit of content; b) access and use of services; and c) preservation of content and sustainability of service. An indicative comparison of challenges and barriers in some major world regions such as Europe, North America and East Asia plus Australia is offered in conclusion.

Two decades of immersion in digital worlds have led to the development of various repository solutions, notably the subject-based repository, research repository, national repository system and institutional repository. However, further development requires a critical appreciation of the current situation as well as an identification of challenges and barriers. In service of further analysis, the main repository solutions are here reconstituted as ideal types. Ideal types are abstract types, derived partly from the history of repositories, partly through logical reasoning. The relevant literature on scholarly communication, open access and repositories is appreciated, though the following is not a literature review but an argument that moves back and forth between abstract ideal types and specific cases.[1] The idea is not to classify each and every repository as belonging unambiguously to a particular type. Rather, the purpose of creating ideal types is to compare and contrast the types so as to generate insight into repository development generally as well as for each individual instance. This implies that the new knowledge thus constituted may enhance the agency of stakeholders and managers in improving and adapting their repository solution.

The four proposed ideal types may be described as follows:
- Subject-based repositories (commercial and non-commercial, single and federated) usually have been set up by community members and are adopted by the wider community. Spontaneous self-archiving is prevalent as the repository is of intrinsic value to scholars. Much of the intrinsic value for authors comes from the opportunity to communicate ideas and results early in the form of working papers and preprints, from which a variety of benefits may result, such as being able to claim priority, testing the value of an idea or result, improving a publication prior to submission, gaining recognition, achieving international attention and so on. As such, subject-based repositories are thematically well defined, and alert services and usage statistics are meaningful for community users;
- Research repositories are usually sponsored by research funding or performing organisations to capture results. This capturing typically requires a deposit mandate. Publications are results, including books, but data may also be considered a result worth capturing, leading to a collection with a variety of items. Because these items constitute a record of science, standards for deposit and preservation must be stringent. The sponsor of the repository is likely to tie reporting functions to the deposit mandate, this being, for example, the reporting of grantees to the funder or the presentation of research results in an annual report. Research repositories are likely to contain high-quality output. This is because its content is peer-reviewed multiple times (e.g. grant application, journal submission, research evaluation) and the production of the results is well funded. Users who are collaborators, competitors or instigating a new research project are most likely to find the collections of relevance;
- National repository systems require coordination - more for a federated system, less for a unified system. National systems are designed to capture scholarly output more generally and not just with a view to preserving a record of scholarship, but also to support, for example, teaching and learning in higher education. Indeed, only a national purpose will justify the national investment. Such systems are likely to display scholarly outputs in the national language, highlight the publications of prominent scholars and develop a system for recording dissertations. One could conceive of such a national system as part of a national research library that serves scholarly communication in the national language and supports public policy, e.g. in generating open educational resources for higher education and enhancing public access to knowledge;
- Institutional repositories contain the various outputs of the institution. While research results are important among these outputs, so are works of qualification, and teaching and learning materials. If the repository captures the whole output, it is both a library and a showcase. It is a library holding an institutional collection, and it is a showcase because the online open access display and availability of the collection may serve to impress and connect, for example, with alumni of the institution or the colleagues of researchers. A repository may also be an instrument of the institution by supporting, for example, internal and external assessment as well as strategic planning. Moreover, an institutional repository could have an important function in regional development. It allows firms, public bodies and civil society organisations to understand immediately what kind of expertise is available locally.

---

[1] Comprehensive and up-to-date bibliographies are available, all edited by Charles W. Bailey, such as Scholarly Electronic Publishing Bibliography, The Open Access Bibliography, and the Institutional Repository Bibliography. The bibliographies may be accessed at http://www.digital-scholarship.org/

Some publication repositories may be identified easily as resembling very much one ideal type rather than another. Some of the classic repositories conventionally identified as subject-based, such as arXiv and RePEc,[2] exhibit few features of another type. Yet, one of the more interesting questions to ask is in how far other elements are present and what this means. ArXiv, for example, is also a research repository, with institutions sponsoring research in high-energy physics being important to its development and success. RePEc, by comparison, has a strong institutional component because the repository is a federated system that relies on input and service from a variety of departments and institutes.

---

[2] http://arxiv.org/; http://repec.org/

To continue with another example, PubMed Central (PMC), at first glance, is a subject-based repository. Acquisition of content, however, only took off once it was declared a research repository capturing the output of publicly funded research (by the NIH). Notably, US Congress passed the deposit mandate, transforming PMC into a national repository. That a parallel, though integrated, repository should emerge in the UK (UK PMC) and Canada (PMC Canada) is thus not surprising. Utilisation of the ideal types outlined above would thus be fruitful in analysing the development of PMC and, presumably, be equally valuable in discussing the future potential of PMC, for example the possible creation of a Europe PMC.[3]

National solutions are increasingly common (and principally may also be regional in form), but vary especially with regard to privileging either research outputs or the institutions. The French HAL system is powered by the CNRS, the most prestigious national research organisation, and thus is strong on making available research results. In Japan, the National Institute of Informatics has supported the Digital Repository Federation, which covers eighty-seven institutions, with mainly librarians working to make the system operational. In Spain, an aggregator and search portal, Recolecta, sits atop a multitude of institutional repositories, with a large variety of items. In Australia, institutional repositories are prominently tied to the national research assessment exercise, with due emphasis on peer reviewed publications.[4]

Repositories have co-evolved with the Internet and thus are characterised by openness, i.e. open source development of the infrastructure and open access to the content. Yet, ongoing expenditure for repositories can only be justified if they are accepted and utilised by scholars – as researchers and lecturers. Repositories thus compete with other channels of publication and data collecting as well as among each other. Therefore, content and service must be combined in an effort to reciprocally enhance their value in supporting scholars in creating new knowledge, be it in the study, laboratory or classroom.

---

[3] http://www.ncbi.nlm.nih.gov/pmc/; http://ukpmc.ac.uk/; http://ukpmc.blogspot.com/2009/07/pmc-canada-will-launch-in-autumn-2009.html; on Europe PMC see Wellcome Library, The Year in Review (2008), p. 13 - http://library.wellcome.ac.uk/assets/wtx055651.pdf; and the Press Release "European research funders throw weight behind UK open access repository" (01 March 2010) - http://www.wellcome.ac.uk/News/Media-office/Press-releases/2010/WTX058744.htm

[4] http://hal.archives-ouvertes.fr/; http://drf.lib.hokudai.ac.jp/drf/; http://www.recolecta.net; for Australia see Kennan, Mary Anne, and Danny A. Kingsley (2009) 'The State of the Nation: A Snapshot of Australian Institutional Repositories', *First Monday* 14(2)

The following inquiry utilizes the distinction between the four ideal types to investigate how repositories may best serve scholarly communication.[5] The rationale is that repositories may have many functions but that unless they serve scholarly communication first and foremost they will not be accepted and used in the long term. Acceptance and usage by the scholarly community is crucial to sustainability. For this, the emphasis must be on identifying challenges and barriers to improved services and asking which types of repositories and what kind of services are needed in future. The argument proceeds as follows. First, two major shifts in digital scholarly communication and their impact on repositories are analysed, namely a) the problem of organising the increasing volume of published knowledge in a fashion that the user is served relevant, interesting and important material; and, b) the need to deliver highly useful services to scholars as authors and readers. Second, challenges and barriers to repository development are discussed in three key dimensions: a) identification and deposit of content; b) access and use of services; and c) preservation of content and sustainability of service. The article closes with an indicative comparison of some major world regions in an effort to help repositories overcome barriers and master the challenges.

### From volume to quality: privileging users

For well over a century, the number of scholars, journals and publications has been increasing steadily, leading to the notion that the volume of published knowledge is doubling in ever-shorter intervals. One response has been increasing specialisation, with users accepting a more limited field of reading. Another response is to be highly selective in reading, for example, by relying on the journal impact factor. However, some stakeholders have been busy developing services for readers to aid them in navigating scientific information according to relevance, interest and quality (e.g. article-level metrics, text mining).

The Internet implies that the volume of published knowledge principally may grow unabated, whereas digitization means that past-published knowledge will be available simultaneously. Moreover, the online dissemination of all kinds of additional material is facilitated, e.g. working papers, conference presentations, data supplements and teaching materials. Further still, it has been possible to finance the increasing volume of published knowledge. While financial constraints may lead to a cap in the volume of knowledge being published, so far technological innovation has brought efficiencies that have allowed the volume to keep growing. In this sense, privileging users in navigating content has become not only more important but also more urgent. What has been the response of repositories?

---

[5] This paper expands and elaborates a comparison of institutional repositories versus more large-scale solutions begun in Romary, Laurent and Chris Armbruster (2010) Beyond Institutional Repositories. *International Journal of Digital Library Systems* 1(1) 44-61; available at SSRN: http://ssrn.com/abstract=1425692

Subject-based repositories have a track record of delivering services to which users interested in specific subject categories (or research fields) may subscribe. Alert services for new papers and impact statistics are delivered by subject, both comprehensively and specifically, meaning that these are more comprehensive than those of publishers, which cover their journal titles only. Insofar as a repository covers one or more subject categories, it may become a one-stop shop, to which publishers would then also be interested to feed details of new publications. By contrast, institutional repositories offer little or no specific services, and users must rely on search-and-find by way of aggregators and generic search engines. Efforts at harvesting or federated search have not, to date, led to the creation of any portal through which a well-defined corpus may be accessed.

National repository systems may offer tuned services as soon as the collection is large enough (and submission rates high enough) for these services to be valuable. For example, the French system HAL offers a subscription service across the disciplines for articles and bibliographic references. Annual submissions passed the mark of fifty thousand in 2008 (for comparison, arXiv also had more than fifty thousand submission in 2008). By contrast, NARCIS, a national portal incorporating the Dutch repository network (DAREnet), does not offer such subscription services (earlier projects, such as the Agricultural Repository News Exchange, did not lead to the adoption of such services). DAREnet originally understood itself as a network of institutional repositories (of the Dutch universities), though NARCIS is now a national portal. HAL is backed by large French research organisations and has a track record of collaborating with the subject-based repositories (i.e. exchanging content).[6] Notably, research repositories have offered services when they are part of (or defined as) a subject-based repository or national system (e.g. CNRS and INRIA within HAL). But if research repositories serve just one institution, there is likely to be neither the critical mass nor a well-defined corpus to merit the launch of any such services.

In sum, subject-based repositories and national systems have provided evidence that they are able to help scholars in navigating large amounts of published knowledge. Institutional repositories have not been able to do so and, given their structural set-up they face difficulties, which could be mitigated if they are aggregated in national systems and cooperate with subject-based portals.[7] Research repositories also could boost subject-based repositories and national systems by delivering high quality content.

---

[6] http://www.narcis.info/; http://hal.archives-ouvertes.fr/

[7] The primary purpose of this article is to compare repositories, and not particularly to criticize institutional repositories, though we note that criticism is growing, e.g. Salo, Dorothea (2008) 'Innkeeper

## Achieving high-value for scholars: dedicated services

Scholarly communication primarily supports the further advancement of knowledge, including the training of the next generation. Repositories are new, and in more than one way, exist in parallel to the existing infrastructure of journals. In this sense, repositories and journals are competing for the attention of readers and authors. It is thus of importance and interest how repositories offer dedicated services.

As subject-based repositories have seen large increases in the number of items available, efforts have gone into securing the quality of submissions. Solutions vary. For example, arXiv requires that any author posting to a research field for the first time is endorsed by established authors. SSRN has increased the number of research fields for which numerous academic editors are selecting items for alert services, while also seeking to enhance its usage and citation metrics.[8] On the whole, subject-based repositories have successfully mastered the challenge of becoming large-scale providers of dedicated services that are relevant and important to scholarly communities, including the best researchers in any field.

By contrast, institutional repositories do not offer dedicated services for any specific community. Rather, they may play an increasingly important role in the assessment of institutions and departments. If so, the institutional policing of a deposit mandate makes sense and repositories may offer assistance to scholars, for example, with librarians and administrators ensuring that submissions are suitable for assessment. Moreover, institutional repositories have begun offering services such as compiling publication lists for scholars and tracking impact, which serve the growing audit culture but also support scholars. Institutional repositories may also focus on supporting scholars in the realm of teaching and learning. If textbooks and course materials (and, possibly, other online tools such as blogs) were tied in, these could be used locally and disseminated worldwide. However, only services from prestigious universities are likely to find a larger number of users (e.g. MIT OpenCourseWare).[9]

---

at the Roach Motel', *Library Trends* 57:2; Basefsky, Stuart (2009) The End of Institutional Repositories and the Beginning of Social Academic Research Service: An Enhanced Role for Libraries. Available at http://www.llrx.com/authors/1133; Albanese, Andrew R. (2009) 'Institutional Repositories: Thinking Beyond the Box', *Library Journal* (March 1); Romary, Laurent and Chris Armbruster (2010) Beyond Institutional Repositories. *International Journal of Digital Library Systems* 1 (1), forthcoming; available at SSRN: http://ssrn.com/abstract=1425692

[8] http://ssrn.com/; http://arxiv.org/
[9] http://ocw.mit.edu/

National repository systems typically offer a multitude of views, by subject or institution, but have developed few services that would match those of subject-based or institutional repositories. Of course, the deployment of any community services requires a well-defined corpus with critical mass. However, given that HAL has reached an annual deposit rate of fifty thousand items, the French system might be a candidate for rolling out more service. Also, the Dutch collection (NARCIS) features more than 180,000 items, the Australian universities boast more than 200,000 records (though only about 30,000 full texts), and the Spanish aggregator (Recolecta) more than 450,000 items (the number of full texts is not clear).[10] This would suggest that more service is possible. More generally, institutional repositories and national repository systems could do more to support scholarly communities, for example, by sharing metadata and content with subject-based services, or aggregating metadata and content in subject-based services. It would be worth exploring whether the RePEc model of institutional feeds for community services could be adopted for other disciplines, particularly for early access to working papers and preprints. Beyond that, national repository systems may have the same function as institutional repositories in research assessment and for teaching and learning.

Research repositories have a (potentially) vital function to play by allowing the tracking of the research frontier by scholars (and other interested parties, e.g. research funders). However, this would require addressing the tension between merely holding a record of science and being an up-to-date communication tool. If research repositories primarily hold final research outputs, defined as peer-reviewed publications, which are only released after formal publication, and possibly only after a publisher's embargo has expired, then this may amount to a public record of science, but the time lapsed between the initially successful funding application (which was judged cutting-edge) and the output available in the repository may be simply too long for the repository to rival subject-based repositories as a community portal. On the other hand, as a research information service this type of repository has value for scholars, for example, with a grant look-up tool, some indication of early results, and measures that allow the tracking of research trends. Notably, UK PMC is developing such services. While UK PMC is also a subject-based repository, these services could be adopted by any generic research repository (serving one or many disciplines).[11] In capturing publications for assessment, research repositories are similar to institutional ones. UK PMC, for example, has developed services to assist authors in deposit (and ease the burden by soliciting publisher deposit) and is planning to develop the user interface in a manner that provides additional tools for research such as seamless access to content across many repositories, text-mining tools and citation tracking.

---

[10] http://www.narcis.info/; Kennan, Mary Anne, and Danny A. Kingsley. "The State of the Nation: A Snapshot of Australian Institutional Repositories." *First Monday* 14, no. 2 (2009); http://www.recolecta.net/buscador/results.jsp

[11] http://ukpmc.ac.uk/grantLookup/; http://ukpmc.ac.uk/ppmc-localhtml/future_plans.html

Providing dedicated services costs money, be it assisted deposit for authors or text-mining technologies. Henceforth, repositories will be worrying about the resources at their disposal. Research repositories may have an advantage because the research funders that sponsor them will have an intrinsic interest to further their development (e.g. PMC and its national instances). National repository systems, if they attract the national government (or an agency on its behalf) as funder, have similar opportunities. Institutional repositories will be very much dependent on the resources of their institution, and this implies that the quality of services will vary widely. The future of subject-based repositories depends on whether they develop a sustainable business model with independent income (e.g. SSRN) or broaden the number of their sponsors (e.g. arXiv).

The above discussion has centred, in a generic manner, on scholarly communities and users as readers and authors. Overall, repository deposit has become systematic on a large scale, and this brings us to discussing the first specific challenges and barriers of repository development.

### Challenges and barriers in identification and deposit

Repositories are available as off-the-shelf software and this has lead many to believe that setting up and networking as many repositories as possible is the right way forward. Yet, the overwhelming majority of repositories already stumble at the first hurdle: identification and deposit of material that is of relevance and interest in scholarly communication. Even in countries with a well-networked repository infrastructure (e.g. United Kingdom, Germany and Australia)[12] deposit rates may be low and the corpus of repositories generally not well defined. By contrast, many subject-based repositories continue to receive voluntarily submissions in large numbers because of a virtuous circle of services and self-archiving, (e.g. arXiv, RePEc, SSRN). National repository systems also attract content (e.g. the French HAL system, the Dutch DAREnet) and may additionally gain content from special initiatives, such as the Dutch project *Cream of Science*, which brought online most of the research papers of the two hundred most important Dutch scientists.

Open access deposit mandates provide an easy way to identify relevant content because they target scholarly output, particularly journal articles, though the emphasis is technically on the author's final peer-reviewed manuscript. A deposit mandate is particularly effective in the case of research funders and prestigious institutions because it may be assumed that the content thus made available is generally not only of relevance but also of high quality, stimulating interest from users. Several observations follow from this:
- Deposit mandates help repositories to identify desirable content, which typically are peer-reviewed publications;
- The mandate asks the scholar to comply, requiring controls, thus distinguishing this type of mandated deposit from self-archiving;
- Institutional repositories may have their character altered insofar as deposit mandates primarily target research results.

---

[12] e.g. http://www.sherpa.ac.uk/; http://www.rsp.ac.uk/; http://www.wrn.aber.ac.uk/en/; http://www.narcis.info; http://www.dini.de/wiss-publizieren/; http://www.arrow.edu.au/;

Deposit continues to be non-trivial. Senior scholars and prolific authors are busy people. Repositories relying on compliance with mandates often find that a system is required for assisted, possibly even automated deposit. For example, UK PMC has found that deposit is eased when publishers deposit directly, with corresponding gains if the deposit version is the final, archival version with full metadata.[13] In the case of open access publishing, deposit by the publisher is usually not a problem. With traditional publishers, the issue is more complicated. Not only do these publishers lobby hard to have an embargo respected, but also the version deposited is not the final one, only the author's final peer reviewed manuscript, without the pagination and editing of the published version. Thus, while deposit mandates bring high quality content, they do so in a form that is not (yet) recognized as authoritative. In this sense, a publisher's repository, with the published version available but subject to an embargo, could be of higher value. A notable example is HighWire Press, a platform on which many journals have embargoes of twelve months before articles become open access (though some journals have an embargo period of only three or six months, and some have a moving wall of several years).[14] To date, more than 1.9 million articles have become available. HighWire Press is preferred by smaller STM publishers, often society publishers, many of which serve the life sciences. A publishers' repository, especially if community-oriented, could be an interesting alternative to repositories that must rely on deposit mandates to acquire content.

Moreover, even for the archiving of the so-called author's final manuscript, assistance may be desirable. For example, direct deposit by publishers would simplify the process. Alternatively, publishers might return a final pre-publication manuscript to the author for deposit. Of course, in this scenario publishers determine an embargo period. Moreover, a service charge might be implemented for returning a controlled version of the final peer-reviewed manuscript. This type of solution should be particularly attractive to institutional repositories that cannot rely on a deposit mandate. It is being tested by the large STM publishers and selected institutional repositories in a joint project (Publishing and the Ecology of European Research – PEER – co-funded by the European Union).[15] A large number of peer-reviewed articles are being made available for open access archiving, with half the articles being directly deposited by publishers and the other half returned to the authors with the permission to self-archive. Our expectation would be that self-archiving will be only sporadic, implying that only assisted deposit, e.g. systematic support from librarians, will ensure that open access deposit occurs systematically.

---

[13] http://ukpmc.ac.uk/ppmc-localhtml/about.html
[14] http://highwire.stanford.edu/about/;

Size, quality and service matter. The large subject-based repositories, as they draw content from top researchers in the field, have few problems with the voluntary deposit of high-quality content. Alternatively, deposit mandates are attractive if implemented by research funders, research organisations (e.g. national academies) and prestigious universities. By contrast, national systems struggle without an articulated strategy that is backed up by a collection policy, subject-based services and, preferably, some mandates from research funders and national organisations. Moreover, institutional repositories would seem dependent on either being integrated with one of the other types of repository (e.g. a national system, feeding subject-based services) or, else, on collaboration with publishers who deposit either the final peer-reviewed manuscript or the final published version. If access is subject to an embargo, however, then a publishers' repository could be the more efficient solution as a one-stop shop with authoritative content.

### Challenges and barriers in access and use

Just because an item has been deposited, it should not be assumed that users find and use it. It is well known that search and find is a problem for repository items because the coverage of specialist search services is limited and in generic search engines repository items will often appear someway down the list. Obviously, the quality and visibility of repository content is crucial to access.

An external measure of access is available through the Ranking Web of World Repositories, performed by the Cybermetrics Lab of the Spanish Research Council (CSIC).[16] The ranking provides a good indicator of the visibility and quality of the repository content (e.g. external inlinks, number of full-text items, scholarly quality). Overall, it is evident that quite a number of the large repositories have difficulties making their content visible, particularly to search engines. Too often, the rich files (texts, data, pictures) are not visible because they lack standard suffixes, for example at CiteSeer X, PubMed Central and RePEc. Beyond that, the large-subject based repositories are ranked highly, as is HAL, as a national system with its institutional domains. As regards the institutional repositories, there is no correlation between the academic prestige of the institution and the repository rank, i.e. most of the prestigious universities and research organisations do not have a repository that is highly ranked (the notable exceptions being, perhaps, the University of California and MIT).

---

[15] http://www.peerproject.eu/

[16] http://repositories.webometrics.info/;

An internal measure of access is available from usage and citation statistics. Usage indicates how often an item has been viewed and downloaded. Usage statistics may be collected by any repository (although true usage for an article might require collation of data from a number of sources, i.e. publisher, repository, any intermediaries or caches).[17] Citation statistics require the tracking of references across the domain, including a decision as to what counts as a citation. However, even repositories with a well-defined corpus (such as arXiv, SSRN and RePEc) have difficulties reporting reliable citation metrics from their corpus alone (no matter whether harvested, federated or centralized). The only available solution at present would be to obtain citation statistics from Google Scholar.[18] The large-subject based repositories have assembled a corpus from which meaningful statistics and rankings may be obtained. For the other types of repositories, however, usage statistics that have any value as information service for scholars are not directly obtainable. This would require international collaboration. For usage, this would need to be an entity trusted by repositories to objectively count and compare usage (e.g. based on Project COUNTER, with usage measured by item).[19] Citation services come into their own, however, only if they are provided across a large and meaningful corpus - only very large subject-based repositories and national systems could have an independent one. To date, only RePEc has meaningful citation statistics.[20] By contrast, publishers have collaborative standards and implemented technology so that users may track and surf on citations (e.g. Crossref, databases like Web of Knowledge, Scopus).[21] For repositories, some of the national repository portals and aggregators (e.g. the Spanish Recolecta, the Bielefeld Academic Search Engine, OAIster)[22] could begin constructing a service that allows citation tracking, surfing and counting.

---

[17] e.g. http://ssrn.com/ (top papers, top authors, top institution)

[18] e.g. http://www.citebase.org/; http://ssrn.com/update/CiteReader.html; http://scholar.google.de/

[19] http://www.projectcounter.org/

[20] Armbruster, Chris (2008) Access, Usage and Citation Metrics: What Function for Digital Libraries and Repositories in Research Evaluation? *Online Currents* 22(5) 168-180; available at http://ssrn.com/author=434782

[21] ibid.

[22] http://www.recolecta.net/buscador/; http://base.ub.uni-bielefeld.de/; http://www.oclc.org/oaister/

Users of repositories increasingly expect access services that go beyond content-for-downloading. Important, for example, is a multiplicity of views on the repository content according to users' interests and the chance to discover related content. These, however, may be quite extensive, such as wanting to view items as according to newness or ranking (overall, in a field) versus wanting an overview of all contributions from an author or research teams, or desiring to assess the output of a department or institute. This need for a multiplicity of views on the repository corresponds to the variety of users, such as the researcher, lecturer, student, peer reviewer, faculty administrator or funder. Another example are advanced users that not only want access but also need the right to mine and re-use the content for the generation of new knowledge. Much of these services require open access 'libre' – legal parameters that enable re-use of scientific information.

Any embargo also has an impact on access and use. The strength of some of the traditional subject-based repositories is that they have attracted early submissions, i.e. working papers and preprints. By contrast, repositories relying on a mandate must respect embargoes, usually of up to six months, but sometimes also longer. In effect, this means that information becomes openly available much later. Time elapses from the pre-print to peer review through to editing and publication. If subsequently an embargo is slapped on, and be it only for six months, then availability is delayed by at least one year, possibly two (counting from the date of first submission through peer review, revisions and production) Also, the online-first function, by which much content becomes available at the publisher's platform before the official publishing date of the journal issue, indirectly lengthens the embargo. Open access publishing has the advantage of providing immediate access upon publication. Insofar as published knowledge is scientific information that powers the search for new knowledge, the timeliness of access matters for usage. It is the circulation of working papers and preprints that, most of all, levels the playing field by making ideas and data available early. If repositories accept embargoes, then the toll-access publication remains the first and premier site for access – and anyone who has no access is disadvantaged as before, because access comes too late. In this scenario, open access publishing would deliver much more value to scholars.

### Challenges and barriers in sustainability and preservation

Sustainability depends ultimately on usage and user satisfaction. The Directory of Open Access Repositories lists more than 1500 repositories, with more than 1100 being institutional repositories.[23] Growth has been significant over the past years, but most research and higher education institutions do not have a repository yet. However, performance capacity is a challenge to any repository. Unlimited high-speed access at any time requires powerful computing facilities, particularly if the repository is to grow and access increases accordingly. Moreover, services for users and authors need to be scaled. Particularly repositories designed as subject-based or national systems face issues of scalability and performance, whereas institutional repositories are limited by design. However, the smaller repositories can therefore also not reap any economies of scale, which indicates that strategy of a repository for each institution is probably the most expensive one overall.

---

[23] http://www.opendoar.org/

Subject-based repositories have demonstrated that they are attractive enough to elicit contributions and support from scholars, including the donation of time and resources, for example, for editorial services. By comparison, institutional repositories struggle already in enticing scholars to deposit. Cost control is important for sustainability, but in the case of the fledgling repository infrastructures it must first be demonstrated that the service is accepted and used by scholars. Of course, institutional repositories may always have a function as a depot for student theses, but the availability of resources for further development would seem to hinge very much on a demonstration of the value to scholarly communication and knowledge production. Only subject-based repositories, and not all of them yet, could claim unambiguously to have demonstrated value and acceptance. National systems and research repositories might seek to side-step the issue by focussing on public access and economic value, and this is important, though it will not legitimate the repository within the scholarly community.

As regards cost control, large repositories may reap economies of scale, but the existing large subject-based repositories have disparate business models, including strong commercial elements and significant volunteer effort. Research repositories and national systems may be run on a small percentage of the overall research budget, which would be justified as the open access dissemination of the research results enables public access and wide impact. Institutional repositories may be independent or part of a division, e.g. the library, but will be largely dependent on the overall financial health of the institution.

If the repository is to have any value over the long term, then a quality control system must be implemented and the integrity of the corpus preserved. One widespread misunderstanding is that repositories are there to archive 'everything' when, in fact, users are ever more concerned about the value and relevance of research results. This is not to say that a search among PhD theses may not lead to interesting results, but it does indicate that the collection and display policy of repositories matter and may influence their acceptance and use. For research repositories this is most easy, as results will have been peer reviewed multiple times. Subject-based repositories typically institute quality controls and markers, such as controlling who may deposit, running series from prestigious institutions and offering statistics that indicate usage. Institutional repositories usually collect items of varying quality and relevance, but by means of a collection policy, research publications could be clearly distinguished, for example, from undergraduate theses.

Long-term preservation and access is an unsolved issue for repositories, but not an urgent one. To date, probably only some subject-based repositories are worth preserving. More generally, national repository systems might count on the national libraries, and research repositories likewise, insofar as their content is recognized as an essential part of the record of science. Of course, efforts at digital preservation are ongoing, and publishers and librarians have, for example, already established CLOCKSS as a joint venture for the preservation of content, particularly when content is no longer available from a publisher.[24] Repositories must attend to the issue of sustainability as a priority, this being primarily about service, usage and cost. The above discussion would suggest that all types of repositories must consider how they can improve services, possibly by collaboration and consolidation to increase the efficiency of provision.

## Comparing the landscape across regions

Particularly in Europe, publicly funded efforts at networking repositories have been substantial, be it in creating national networks (Netherlands, UK) or in fostering the development of European infrastructure, standards and a community of practice (DRIVER I & II).[25] A good overview of the European landscape is provided by the DRIVER search portal, through the 'browse by repository' function.[26] This reveals that most repositories continue to hold very limited content already when it comes to records, let alone full text access. This is unsurprising if one considers the low number of institutional open access mandates in Europe. Much more consequential are the research funder mandates, including those at the European level.[27] For most of Europe a disjoint must be diagnosed between policy and infrastructure. The infrastructure is being built for institutional deposit, but at the policy level research funder mandates are more numerous and efficacious. The question Europe needs to address is whether it can repurpose all those repositories that are viewed primarily as institutional into ones that serve to capture research outputs. Notably, most of the national networks feature university repositories, often of the more prominent or all universities – as opposed to all and any higher education institutions. This could imply that a shift is possible to the ideal type of the research repository as guiding vision.

---

[24] http://www.clockss.org/

[25] A good summary is Vernooy-Gerritsen, Marjan, Gera Pronk, and Maurits van der Graaf (2009) 'Three Perspectives on the Evolving Infrastructure of Institutional Research Repositories in Europe', *Ariadne* 59 (April)

[26] http://search.driver.research-infrastructures.eu/

[27] ROARMAP at http://www.eprints.org/openaccess/policysignup/

According to the Webometrics Ranking of Repositories, the one repository system that is successful is the French one.[28] Yet, most European countries are not following this lead. Notable about the French system is that the national research organisations have organized the infrastructure and contribute most of the content. Insofar as other European players shift focus to capturing research output, a common ground could emerge. The optimal mix might be strong national systems that organize the deposit and preservation, while services become community specific – thus becoming compatible with the subject-based repositories.

North America, and the United States in particular, are home to most of the successful subject-based repositories (e.g. arXiv, CiteSeer, PMC, RePEc, Smithsonian NASA ADS, SSRN), some notable institutional exceptions notwithstanding (Illinois, California, MIT). Europe has missed out on headquartering the repositories (e.g. RePEc was conceived in Europe), but is actively participating in all, sometimes by building auxiliary services (e.g. UK PMC, Europe PMC). Given how US institutions dominate the university rankings, it is notable that this is not so for repository rankings. One may adduce that the strong performance of the subject-based repositories means that institutional solutions are more attractive for an open access deposit mandate (e.g. Harvard) or another specific aim, such as building a digital library (e.g. California), and enhancing teaching and learning worldwide (e.g. MIT OCW).

Europe and the United States have been the main players. The above analysis indicates that between these two, a solution could be found that could be scalable globally. The two components would be subject-based repositories and systems of research repositories, the later variously being national systems (e.g. France, Netherlands), institutional systems (e.g. University of California, Max Planck Society), and possibly single prestigious research institutions (e.g. Harvard). The focus on capturing research results and supporting scholarly communities through services could lead, finally, to widespread acceptance and use of repositories in scholarly communication, which is vital to future sustainability.

At first glance, such a solution would seem extendable to other players in the repository landscape as defined by a good ranking, i.e. institutions in Australia, Canada and Japan.[29] To all intent and purpose, Japan and

---

[28] http://repositories.webometrics.info/

[29] In the Webometrics ranking of July 2009, the first Brazilian repository ranks 57 (University of Sao Paolo), the first Chinese at 88 (National Tsing Hua University), and the first Indian at 121 (Indian Institute of Science Bangalore).

Australia are building a national systems based on capturing the research output of universities. Canada is pursuing the dual strategy of joining subject-based repositories (e.g. PMC Canada) and supporting institutional repositories, and more than four out of five members of the Canadian Association of Research Libraries have a repository.[30] All in all, this indicates that the global solution proposed above would be concomitant with national systems, including their function of making content available in the national language(s).

We may conclude that it is principally possible to build a repository infrastructure that scales globally and is of value in scholarly communication, but we must also recognize that the above discussion suggests that some path breaking change is required, particularly in upgrading and repurposing institutional repositories, and in the collaboration among repositories and with service providers to enhance access and usage. Whether this will happen remains to be seen, but in any case the ideal types developed and the subsequent discussion may be used to review and discuss developments in future.

---

[30] http://www.carl-abrc.ca/projects/institutional_repositories/institutional_repositories-e.html